\documentclass[10pt]{iopart}

\usepackage{iopams}  

\def\Journal#1#2#3#4{#4 {#1} {\bf #2} #3}

\def\EPJC{Eur. Phys. J. C}

\def\JMP{J. Math. Phys.}

\def\NPBOLD{Nucl. Phys.}

\def\PPNP{Prog. Part. Nucl. Phys.}
\def\PLBOLD{Phys. Lett.}

\def\RPP{Rept. Prog. Phys.}
\def\PRL{Phys. Rev. Lett.}
\def\PRD{Phys. Rev. D}

\def\PR{Phys. Rev.}
\def\PTP{Prog. Theor. Phys.}
\def\PTPSUPPL{Prog. Theor. Phys. Suppl.}

\def\RMP{Rev. Mod. Phys.}
\def\RMATHP{Rev. Math. Phys.}

\begin{document}

\title[Spinor Representation of $O(3)$ for $S_4$]{Spinor Representation of $O(3)$ for $S_4$}

\author{Teruyuki Kitabayashi$^1$, Masaki Yasu\`{e}$^2$}

\address{Department of Physics, Tokai University,4-1-1 Kitakaname, Hiratsuka, Kanagawa 259-1292, Japan}
\ead{$^1$ teruyuki@tokai-u.jp, $^2$ yasue@keyaki.cc.u-tokai.ac.jp}
\vspace{10pt}
\begin{indented}
\item[] \today 
\end{indented}

\begin{abstract}
All possible permutations in the discrete $S_4$ group are classified by three rotation angles associated with the orthogonal group $O(3)$. We construct a spinor representation ${\bf 2}_D$ of $O(3)$, which is transformed by three 4$\times$4 matrices corresponding to three Pauli matrices in $SO(3)$. An irreducible decomposition of ${\bf 2}_D \otimes {\bf 2}_D$ supplies a vector representation of {\bf 3} of $O(3)$, thereby, of $S_4$. Our construction is consistent with the mathematical fact that $O(3)=SO(3)\times \boldsymbol{Z}_2$. The $\boldsymbol{Z}_2$ parity in the spinorial space is described by a block off-diagonal matrix as the spinorial parity operator, whose eigenvalues are $\pm 1$ consistent with $\boldsymbol{Z}_2$.
\end{abstract}

%
%
%
%
%

\section{Introduction}
Symmetries have taken a significant r\^{o}le in particle physics since the discovery of the $SU(3)$ symmetry in the hadron physics \cite{Sakata1956PTP,Ogawa1956PTP,Ikeda1956PTP,Yamaguchi1956PTPSUPPL,GelMann1961PR,Neeman1961NPBOLD,GelMann1964PLB,Zweig1964CERN}. Nowadays, a possible r\^{o}le of symmetries is more important to discuss the current issues associated with neutrino oscillations, which contain the problem of neutrino masses. The origin of neutrino masses and their mass hierarchy inducing neutrino mixings, which are experimentally confirmed \cite{Capozzi2018PPNP,TORTOLA2018Nu2018}, is not understood in the standard model and is expected to well understood by enlarging the standard model symmetry to include massive neutrinos. Modified standard models involve new flavor symmetries under which different species of neutrinos carry different flavor charges \cite{Everett2018Nu2018}.

As a flavor symmetry of neutrinos, various non-Abelian discrete symmetries have been discussed to determine a mass spectrum of neutrinos as well as their mixings \cite{Ishimori2010PTP,Altarelli2010RMP,King-Luhn2013RPP}. Among others, let us focus our attention to the discrete $S_4$ symmetry \cite{Lam2008PRL}. It is known that $S_4$ deals with all permutations among four objects $\{x_1,x_2,x_3,x_4\}$. Since $\xi_0 \propto x_1+x_2+x_3+x_4$ is obviously invariant under any permutation of $S_4$, $\xi_0$ serves as a singlet ${\bf 1}$ of $S_4$. The remaining three degrees of freedom are described by a column vector $\boldsymbol{\xi}$:
\begin{eqnarray}
\boldsymbol{\xi}=\left(
  \begin{array}{c}
     \xi_1   \\
     \xi_2  \\
     \xi_3  \\
  \end{array}
  \right),
\label{Eq:threexi}
\end{eqnarray}
where $\xi$'s are orthogonal to each other and also to $\xi_0$, and forms a triplet ${\bf 3}$ of $S_4$.  All elements of $S_4$ acting on {\bf 3} are, therefore, represented by $3 \times 3$ rotation matrices in the orthogonal group $O(3)$, which allows any dimensional representations other than three dimensional vector representation \cite{Etesi1998JMP,King2018}.

In this paper, we construct a spinor representation of $O(3)$ using two-component Pauli spinors of the special orthogonal group $SO(3)$, which turns out to be a four-component spinor to include a reflection in $O(3)$. This spinor is transformed by $4 \times 4$ complex rotation matrices in $O(3)$. Our four-component spinor is essentially two-component one representing a doublet of $O(3)$, which is denoted by ${\bf 2}_D$. As a result, $S_4$ will contain two classes of doublets: one is the conventional one produced by ${\bf 3} \otimes {\bf 3}$ and the other is a new one generating {\bf 3} out of ${\bf 2}_D \otimes {\bf 2}_D$. 

Mathematically speaking, there exists a double covering of $O(n)$, which is known as a $Pin(n)$ group and its subgroup is $Spin(n)$, which is a double covering of $SO(n)$. Some of physics-oriented studies are found in Ref.\cite{Trautman1994,DeWitt-Morette1990PRD,Berg2001RMATHP,Janssens2017}. The spinor of $O(3)$ can be treated by the $Pin(3)$ group. However, we would like to develop more explicit and practical discussions on the spinor of $O(3)$ so that its applicability in particle physics becomes more visible. In fact, an explicit construction of the $O(3)$ spinor enables us to know how to assign particles to the four-component spinor.

Our main aim is to shed light on the use of a spinor representation in $S_4$. The other aim is to show an example of our spinor of $O(3)$ participating in particle physics and cosmology. This paper is organized as follows. In Sec. \ref{sec:vector_3dim}, we show a brief review of triplet representation ${\bf 3}$ in a vector space. In Sec.\ref{sec:spinor_2dim}, we construct a spinor representation as a doublet representation ${\bf 2}_D$. To get a hint on a spinor representation of $O(3)$, we examine a \lq\lq spinor\rq\rq representation of $S_3$ associated with $O(2)$. In Sec.\ref{sec:parityoperator}, the $\boldsymbol{Z}_2$ parity in the decomposition of $O(3)$ as $O(3)=SO(3)\times \boldsymbol{Z}_2$ is explained by the existence of an off-diagonal matrix as a spinor parity operator, whose eigenvalues are $\pm 1$ indicating $\boldsymbol{Z}_2$. The spinor parity operator is found to be responsible for  the phase transformation of the $O(3)$ spinor induced by that of the $SO(3)$ spinor. Briefly described in Sec.\ref{sec:scenario} is an example of application of the spinor of $O(3)$ to particle physics and cosmology. Section \ref{sec:summary} is devoted to a summary. 

\section{Vector representation}\label{sec:vector_3dim}
All permutations among four objects $\{x_1,x_2,x_3,x_4\}$ to the permuted state $\{x_1',x_2',x_3',x_4'\}$ can be described by  
\begin{eqnarray}
x^\prime= U x,
\end{eqnarray}
where
\begin{eqnarray}
x'=\left(
  \begin{array}{c}
     x_1'   \\
     x_2'  \\
     x_3'   \\
     x_4'   \\
  \end{array}
\right),
\quad
x=\left(
  \begin{array}{c}
     x_1   \\
     x_2  \\
     x_3   \\
     x_4   \\
  \end{array}
\right), 
\end{eqnarray}
and $U$ denotes a $4 \times 4$ matrix. The induced four objects of $\xi_{0,1,2,3}$ in Eq.(\ref{Eq:threexi}) form $\xi$:
\begin{eqnarray}
\xi  =
\left(
  \begin{array}{c}
\xi_0 \\
\boldsymbol{\xi}
  \end{array}
  \right),
\label{Eq:xi}
 \end{eqnarray}
which is also transformed into $\xi^\prime$ by another $U$, $U^\prime$, as
\begin{eqnarray}  
\xi' = U' \xi.
\end{eqnarray}
Since $\xi$ is related to $x$ by an appropriate $4 \times 4$ matrix $W$: 
\begin{eqnarray}
\xi=Wx,
\end{eqnarray}
$U^\prime$ is given by
\begin{eqnarray}  
U' = WUW^{-1}.
\label{Eq:UprimeGeneral}
\end{eqnarray}
Because of $\xi_0^\prime = \xi_0$, a transformation $\xi \rightarrow \xi'$ is represented by a block diagonal matrix
\begin{eqnarray}
\left(
  \begin{array}{c}
     \xi'_0  \\
     \xi'_1   \\
     \xi'_2  \\
     \xi'_3  \\
  \end{array}
  \right)
  &=&
\left(
\begin{array}{cccc}
1 & 0 & 0 & 0 \\
0 &* & * & *  \\
0 &* & * & * \\
0 & * & * & * \\
\end{array}
\right)
  \left(
  \begin{array}{c}
     \xi_0  \\
     \xi_1   \\
     \xi_2  \\
     \xi_3  \\
  \end{array}
  \right),
\end{eqnarray}
and $\xi_{1,2,3}$ are mixed by three-dimensional rotations.

Since rotations in three dimensions are described in terms of three angles $\theta_{12}$, $\theta_{23}$ and $\theta_{13}$, we define the matrix $U'$ in Eq.(\ref{Eq:UprimeGeneral}) to be:
\begin{eqnarray}
U^\prime(\sigma,\theta_{12},\theta_{13},\theta_{23}) =  \left(
  \begin{array}{cc}
     1 & {\bf 0}    \\
     {\bf 0} & T\left(\sigma,\theta_{12},\theta_{13},\theta_{23}\right)    \\
  \end{array}
  \right), 
\label{Eq:Uprime}
\end{eqnarray}
where $\sigma={\rm det}(T)(=\pm 1)$ takes care of the reflection, and
\begin{eqnarray}
&&T(\sigma,\theta_{12},\theta_{13},\theta_{23}) \nonumber \\
&& \quad =
  \left(
  \begin{array}{ccc}
    1 & 0 & 0\\
    0 & \cos \theta_{23} & -\sin \theta_{23}  \\
    0 & \sin \theta_{23}  & \cos\theta_{23}   \\
  \end{array}
  \right) 
   \left(
  \begin{array}{ccc}
    \cos \theta_{13} & 0 &  \sin \theta_{13}  \\
    0 & 1 & 0\\
    -\sin \theta_{13}  & 0 & \cos\theta_{13}   \\
  \end{array}
  \right) \nonumber \\
  &&\qquad \times
 \left(
  \begin{array}{ccc}
    \sigma \cos \theta_{12} & - \sin \theta_{12} & 0 \\
    \sigma \sin \theta_{12}  & \cos\theta_{12} & 0  \\
    0 & 0 & 1\\
  \end{array}
  \right).
  \end{eqnarray}
This $3 \times 3$ matrix $T$ acts on the triplet representation ${\bf 3}$ of $S_4$ in a vector space. Considering the arbitrariness of the inclusion of $\sigma$ in $T$, we may define
\begin{eqnarray}
T_{23}(\sigma,\theta)
& = &
  \left(
  \begin{array}{ccc}
    1 & 0 & 0\\
    0 & \sigma \cos \theta & -\sin \theta  \\
    0 & \sigma \sin \theta  & \cos\theta   \\
  \end{array}
  \right),
    \nonumber \\
T_{31}(\sigma,\theta)
& = &
   \left(
  \begin{array}{ccc}
    \cos \theta & 0 &  \sigma \sin \theta  \\
    0 & 1 & 0\\
    -\sin \theta  & 0 & \sigma  \cos\theta  \\
  \end{array}
  \right),
\label{Eq:T23T13T12}\\
T_{12}(\sigma,\theta)
& = &
 \left(
  \begin{array}{ccc}
    \sigma \cos \theta & - \sin \theta & 0 \\
    \sigma \sin \theta  & \cos\theta & 0  \\
    0 & 0 & 1\\
  \end{array}
  \right).
\nonumber
  \end{eqnarray}
A rotation of the vector $\vec{\xi}$ (=$\xi_1 \vec{i} + \xi_2 \vec{j} + \xi_3 \vec{k}$) in the $(i,j)$-plane is generated by
\begin{equation}
\boldsymbol{\xi}^\prime = T_{ij}(\sigma,\theta)\boldsymbol{\xi},
 \end{equation}
where $(i,j)$ =(1,2), (2,3) and (3,1) (hereafter, we use $ij$ without referring the numbers). Any rotations in three dimensions can be generated by
\begin{eqnarray}
T=T_{23}T_{31}T_{12}.
\label{Eq:T=T23T13T12}
 \end{eqnarray}

It is noted that the matrices $T_{12,23,31}(\sigma,\theta)$ satisfy the relations of 
\begin{eqnarray}
T_{ij}(1, \theta_1)T_{ij}(1, \theta_2) &=& T_{ij}(1, \theta_1+\theta_2), \nonumber \\
T_{ij}(1, \theta_1)T_{ij}(-1, \theta_2) &=& T_{ij}(-1, \theta_1+\theta_2), \nonumber \\
T_{ij}(-1, \theta_1)T_{ij}(1, \theta_2) &=& T_{ij}(-1, \theta_1-\theta_2), \nonumber \\
T_{ij}(-1, \theta_1)T_{ij}(-1, \theta_2) &=& T_{ij}(1, \theta_1-\theta_2).
\label{Eq:relationsOfTij}
\end{eqnarray}
The tensor products of $S_4$ in a vector space are given by
\begin{eqnarray}
{\bf 3} \otimes {\bf 3} &=& {\bf 3'} \otimes {\bf 3'}={\bf 1} \oplus {\bf 3} \oplus {\bf 2}\oplus {\bf 3'}, \nonumber \\
{\bf 3} \otimes {\bf 3'} &=& {\bf 1'} \oplus {\bf 3'} \oplus {\bf 2}\oplus {\bf 3}, \nonumber \\
{\bf 3} \otimes {\bf 2} &=& {\bf 3'} \otimes {\bf 2} = {\bf 3} \oplus {\bf 3'}, \nonumber \\
{\bf 2} \otimes {\bf 2} &=& {\bf 1} \oplus {\bf 2} \oplus {\bf 1'},
\end{eqnarray}
together with the obvious products of ${\bf 1} \otimes {\bf 1} = {\bf 1^\prime} \otimes {\bf 1^\prime} = {\bf 1}$, ${\bf 1} \otimes {\bf 1^\prime} = {\bf 1'}$, ${\bf 1} \otimes {\bf 3} = {\bf 1'} \otimes {\bf 3'} = {\bf 3}$, ${\bf 1} \otimes {\bf 3'} = {\bf 1'} \otimes {\bf 3} = {\bf 3'}$ and ${\bf 1} \otimes {\bf 2} = {\bf 1'} \otimes {\bf 2} = {\bf 2}$, where prime stands for an antisymmetric representation \cite{Itzykson1966RMP,Ishimori2010PTP}.
\section{Spinor representation}\label{sec:spinor_2dim}
For $SO(3)$, the rotation matrices are restricted to the case of $\sigma=1$.  It is well known that the spinor representation uses the three Pauli matrices, $\tau_{1,2,3}$, which yield 
\begin{eqnarray}
&&S_{23}(1,\theta)=\exp(-i\frac{\tau_1}{2}\theta),
\nonumber \\ 
&&S_{31}(1,\theta)=\exp(-i\frac{\tau_2}{2}\theta),
\label{S233112}\\
&&S_{12}(1,\theta)=\exp(-i\frac{\tau_3}{2}\theta),
\nonumber
\end{eqnarray}
respectively, corresponding to $T_{23,31,12}$.  For a two-component spinor of $SO(3)$, 
\begin{eqnarray}
\alpha=
\left(
  \begin{array}{c}
\alpha_1 \\
\alpha_2
  \end{array}
  \right),
\end{eqnarray}
a vector $\vec{t}$ can be defined by 
\begin{equation}
\vec{t} = \alpha^\dag\vec{\tau}\alpha
\label{Eq:vector-t},
\end{equation}
where
\begin{equation}
\vec{\tau}= \tau_1 \vec{i} + \tau_2 \vec{j} + \tau_3 \vec{k}.
\end{equation}
As well known, under the transformation of 
\begin{equation}
\alpha^\prime = S_{ij}(1,\theta)\alpha,
\end{equation}
the vector $\vec{t}$ is transformed into $\vec{t}^\prime$ according to
\begin{equation}
\boldsymbol{t}^\prime = T_{ij}(1,\theta)\boldsymbol{t}
\end{equation}
where $\boldsymbol{t} = (t_1,t_2, t_3)^T$.

To get a hint to include the reflection for a spinor representation of $O(3)$, let us consider the simplest case of $O(2)$. For the vector representation, the rotation matrix $T\left( {\sigma ,\theta } \right)$ is given by
\begin{eqnarray}
{T\left( {\sigma ,\theta } \right) = \left( 
{\begin{array}{*{20}{c}}
{\sigma \cos \theta }&{ - \sin \theta }\\
{\sigma \sin \theta }&{\cos \theta }
\end{array}} \right)},
\label{SO2vector}
\end{eqnarray}
acting on $(x_1,x_2)$, which is transformed into $(x^\prime_1,x^\prime_2)$.  As a \lq\lq spinor\rq\rq  representation, $z = x_1+ix_2$ is transformed into $z^\prime=e^{i\theta}z$ ($\sigma=1$) and into $z^\prime=-e^{i\theta}z^\ast$ ($\sigma=-1$), which suggest the use of 
\begin{eqnarray}
\boldsymbol{z}  =
\left(
  \begin{array}{c}
z \\
z^\ast
  \end{array}
  \right),
\label{Eq:SO2z}
 \end{eqnarray}
as a new basis. We obtain that $\boldsymbol{z}^\prime = S(\sigma,\theta)\boldsymbol{z}$ with the following $S(\sigma,\theta)$:
\begin{eqnarray}
&&S\left( {1 ,\theta } \right) = \left( {\begin{array}{*{20}{c}}
{{e^{i\theta }}}&0\\
0&{{e^{ - i\theta }}}
\end{array}} \right),
\nonumber \\
&&S\left( {-1 ,\theta } \right) =  - \left( {\begin{array}{*{20}{c}}
0&{{e^{i\theta }}}\\
{{e^{ - i\theta }}}&0
\end{array}} \right).
\label{SO2spinor}
\end{eqnarray}
Since it can be further found that Eq.(\ref{SO2spinor}) is transformed into Eq.(\ref{SO2vector}), nothing new arises from Eq.(\ref{SO2spinor}).

Although nothing new has been found in the case of $O(2)$, it suggests that spinor representation matrices of $O(3)$ may take the similar forms of 
\begin{eqnarray}
\left(
  \begin{array}{cc}
     a& 0  \\
     0& a^\ast  \\
  \end{array}
\right),
\qquad
\left(
  \begin{array}{cc}
    0& a  \\
     a^\ast& 0  \\
  \end{array}
\right).
\label{Eq:PossibleForm}
\end{eqnarray}
Since the entry of $a$ can be one of the Pauli matrices, we expect that representation matrices are $4 \times 4$ matrices to take care of the reflection in $O(3)$. Furthermore, this construction implies that a four-component spinor is something like $(\alpha, \alpha^\ast)$. 

We introduce a four-component spinor $\Phi$ defined by
\begin{eqnarray}
\Phi=\left(
  \begin{array}{c}
     \alpha   \\
     \alpha^\ast  \\
  \end{array}
\right),
\end{eqnarray}
as a doublet representation ${\bf 2}_D$.  The extended representation matrix of $S_{ij}(1,\theta)$, now acting on $\Phi$, can be assumed to be the following form as suggested by the first form of Eq.(\ref{Eq:PossibleForm}):
\begin{eqnarray}
D_{ij}(1,\theta)=\left(
  \begin{array}{cc}
     S_{ij}(1,\theta) & 0 \\
     0 & S_{ij}^\ast (1,\theta)  \\
  \end{array}
\right),
\end{eqnarray}
which transforms the spinor $\Phi$ into $\Phi^\prime$ as follows:
\begin{equation}
\Phi^\prime= D_{ij}(1,\theta)\Phi.
\end{equation}

The extension of the Pauli matrices is straightforward and simply replaces $\tau_i$ ($i=1,2,3$) with the following $4 \times 4$ matrix $\Gamma_i$:
\begin{eqnarray}
\Gamma_i=\left(
  \begin{array}{cc}
     \tau_i & 0   \\
     0  & \tau^\ast_i\\
  \end{array}
\right).
\end{eqnarray}
A vector $\vec{s}$ can be defined by
\begin{eqnarray}
\vec{s}=\Phi^\dag \vec{\Gamma}\Phi,
\end{eqnarray}
where 
\begin{equation}
\vec{\Gamma}=\Gamma_1 \vec{i} + \Gamma_2 \vec{j} + \Gamma_3 \vec{k}.
\end{equation}
This vector $\vec{s}$ is transformed under $\Phi \rightarrow \Phi^\prime = D_{ij}(1,\theta)\Phi$ as
\begin{eqnarray}
\vec{s^\prime}&=&\Phi^{\prime\dag}\vec{\Gamma}\Phi^\prime \nonumber \\
&=& \Phi^\dag \left(
  \begin{array}{cc}
     u^\dag(\theta) \vec{\tau} u(\theta) & 0   \\
     0  & u^T(\theta) \vec{\tau}^\ast  u^*(\theta) \\
  \end{array}
\right) \Phi 
\nonumber \\
&\equiv& s_1^\prime \vec{i} + s_2^\prime \vec{j} + s_3^\prime \vec{k},
\end{eqnarray}
where $u(\theta)$ is either one of $S_{ij}(1,\theta)$. It is readily found that the resulting transformation property of $\vec{s}$ is the same as $\vec{t}$ of Eq.(\ref{Eq:vector-t}), therefore, as $\vec{\xi}$.

Let us proceed to construct appropriate matrices for the $\sigma=-1$ group of $S_{ij}(-1,\theta)$. The constraints arise from the required transformation property given by $T_{ij}(-1,\theta)$ acting on $\boldsymbol{\xi}$. To find explicit forms of $S_{ij}(-1,\theta)$, it is useful to employ the conventional definition of complex numbers arranged from a vector $\vec{\xi}$: 
\begin{eqnarray}
{\boldsymbol{Z} \equiv \left( {\begin{array}{*{20}{c}}
{{\xi_3}}&{{\xi_1} - i{\xi_2}}\\
{{\xi_1} + i{\xi_2}}&{ - {\xi_3}}
\end{array}} \right)},
\end{eqnarray}
from which Eq.(\ref{S233112}) can be derived by requiring that 
\begin{equation}
\boldsymbol{Z}^\prime = u(\theta)\boldsymbol{Z}{u(\theta)^{ - 1}}
\label{ZtoZprime}
\end{equation}
induced by Eq.(\ref{Eq:T23T13T12}).  The reflection can be implemented in the transformation of $\boldsymbol{Z}^\ast$ into $\boldsymbol{Z}^\prime$ as suggested by $z^\ast$ transformed into $z^\prime$ in the case of $O(2)$:
\begin{equation}
\boldsymbol{Z}^\prime = v(\theta)\boldsymbol{Z}^\ast{v(\theta)^{ - 1}},
\label{ZtoZastprime}
\end{equation}
where $v(\theta)$ is a $2 \times 2$ matrix and is determined, up to possible phases, so as to reproduce $\boldsymbol{\xi}^\prime = T_{ij}(-1,\theta)\boldsymbol{\xi}$. Again, $\boldsymbol{Z}$ and $\boldsymbol{Z}^\ast$ are grouped into a block diagonal $4 \times 4$ matrix of ${\rm diag.}(\boldsymbol{Z},\boldsymbol{Z}^\ast)$, which enables us to treat Eqs.(\ref{ZtoZprime}) and (\ref{ZtoZastprime}) in a unified way. 

If $\Phi$ is transformed by a matrix $D_{ij}(-1,\theta)$:
\begin{equation}
\Phi^\prime = D_{ij}(-1,\theta)\Phi,
\label{Eq:DijPsi}
\end{equation}
as a result of the action $T_{ij}(-1,\theta)$, $D_{ij}(-1,\theta)$ turns out to take the second form of Eq.(\ref{Eq:PossibleForm}) with $a=v(\theta)$.
Although, the computation is a bit involved, the forms of $D_{ij}(-1,\theta)$ are very simple and described by $S_{ij}(1,\theta)$ of $SO(3)$. The obtained candidates of $D_{ij}(-1,\theta)$ are as follows:
\begin{equation}
D_{ij}(-1, \theta)
=
\left(
  \begin{array}{cc}
     0&  S_{ij}(-1,\theta)  \\
     S^\ast_{ij}(-1,\theta)&0  \\
  \end{array}
\right),\\
\label{Eq:DijO3}
\end{equation}
where
\begin{eqnarray}
S_{23}(-1,\theta)&=& S_{23}(1,\theta),
\nonumber \\
S_{31}(-1,\theta)&=& -iS_{31}(1,\theta)\tau_1,
\label{Eq:SijO3}\\
S_{12}(-1,\theta)&=& iS_{12}(1,\theta)\tau_3.
\nonumber
\end{eqnarray}
and the phases are so determined to be consistent with $O(3)=SO(3)\times \boldsymbol{Z}_2$ (See Sec.\ref{sec:parityoperator}).

To be consistent, we should confirm that
\begin{equation}
\boldsymbol{s}^\prime = T_{ij}(-1,\theta)\boldsymbol{s},
\label{Eq:SijTij}	
\end{equation}
where $\boldsymbol{s}$ (=$(s_1,s_2,s_3)^T$), is correctly derived by the action of Eq.(\ref{Eq:DijPsi}). The vector $\vec{s^\prime}$, which is generated by $D_{ij}(-1,\theta)$, is described by
\begin{eqnarray}
\vec{s^\prime}= \Phi^\dag \left(
  \begin{array}{cc}
     v^T(\theta) \vec{\tau}^\ast  v^\ast(\theta) & 0   \\
     0  & v^\dag(\theta) \vec{\tau} v(\theta) \\
  \end{array}
\right) \Phi, 
\label{Eq:induced}
\end{eqnarray}
where $v(\theta)$ is either one of $S_{ij}(-1,\theta)$. The derivations make the full use of $S_{ij}(1,\theta)$ of $SO(3)$:

\begin{itemize}
\item For $T_{23}(-1,\theta)$, 
\begin{eqnarray}
&&S_{23}^T\left( { - 1,\theta } \right){{\vec \tau }^\ast }S_{23}^\ast \left( { - 1,\theta } \right) 
\nonumber \\
&&
\quad = S_{23}^\dag \left( {1, - \theta } \right)( {{\tau _1}\vec{i} - {\tau _2}\vec{j} + {\tau _3}\vec{k}}){S_{23}}\left( {1, - \theta } \right)
\nonumber \\
&&
\quad = {\tau _1}\vec{i} + \left( {{\tau _2}\cos \left( { - \theta } \right) - {\tau _3}\sin \left( { - \theta } \right)} \right)(  - \vec{j} )
\nonumber \\
&&
\quad + \left( {{\tau _2}\sin \left( { - \theta } \right) + {\tau _3}\cos \left( { - \theta } \right)} \right)\vec{k},
\end{eqnarray}
which yields $\boldsymbol{s^\prime}=T_{23}(-1,\theta)\boldsymbol{s}$;
\item For $T_{31}(-1,\theta)$, 
\begin{eqnarray}
&&S_{31}^T\left( { - 1,\theta } \right){{\vec \tau }^ * }S_{31}^ * \left( { - 1,\theta } \right)
\nonumber \\
&&
\quad = S_{31}^\dag \left( {1, - \theta } \right)( {{\tau _1}\vec{i} + {\tau _2}\vec{j} - {\tau _3}\vec{k}} ){S_{31}}\left( {1, - \theta } \right)
\nonumber \\
&&
\quad = \left( {{\tau _1}\cos \left( { - \theta } \right) + {\tau _3}\sin \left( { - \theta } \right)} \right)\vec{i}
\nonumber \\
&&
\quad + {\tau _2}\vec{j} + \left( { - {\tau _1}\sin \left( { - \theta } \right) + {\tau _3}\cos \left( { - \theta } \right)} \right)( { - \vec{k}}),
\end{eqnarray}
which yields $\boldsymbol{s^\prime}=T_{31}(-1,\theta)\boldsymbol{s}$;
\item For $T_{12}(-1,\theta)$, 
\begin{eqnarray}
&&
S_{12}^T\left( { - 1,\theta } \right){{\vec \tau }^ * }S_{12}^ * \left( { - 1,\theta } \right)
\nonumber \\
&&
\quad  = S_{12}^\dag \left( {1, - \theta } \right)( { - {\tau _1}\vec{i} + {\tau _2}\vec{j} + {\tau _3}\vec{k}} ){S_{12}}\left( {1, - \theta } \right)
\nonumber \\
&&
\quad  = \left( {{\tau _1}\cos \left( { - \theta } \right) - {\tau _2}\sin \left( { - \theta } \right)} \right)( { - \vec{i}})
\nonumber \\
&&
\quad  + \left( {{\tau _1}\sin \left( { - \theta } \right) + {\tau _2}\cos \left( { - \theta } \right)} \right)\vec{j} + {\tau _3}\vec{k},
\end{eqnarray}
which yields $\boldsymbol{s^\prime}=T_{12}(-1,\theta)\boldsymbol{s}$.
\end{itemize}
As a result, we find that Eq.(\ref{Eq:SijTij}) is reproduced. Finally, it should be noted that the relations
\begin{eqnarray}
D_{ij}(1, \theta_1)D_{ij}(1, \theta_2) &=& D_{ij}(1, \theta_1+\theta_2), \nonumber \\
D_{ij}(1, \theta_1)D_{ij}(-1, \theta_2) &=& D_{ij}(-1, \theta_1+\theta_2), \nonumber \\
D_{ij}(-1, \theta_1)D_{ij}(1, \theta_2) &=& D_{ij}(-1, \theta_1-\theta_2), \nonumber \\
D_{ij}(-1, \theta_1)D_{ij}(-1, \theta_2) &=& D_{ij}(1, \theta_1-\theta_2),
\end{eqnarray}
are satisfied for ${\bf 2}_D$, which are the same as those for {\bf 3} as shown in Eq.(\ref{Eq:relationsOfTij}).

It is certain that our choice of $D_{ij}(\pm 1,\theta)$ supplies correct representation matrices for the spinor of $\Phi$ as ${\bf 2}_D$. The reflection on the $SO(3)$ spinor causes the transformation owing to, for example, the $2$-$3$ rotation as follows:
\begin{equation}
\alpha^\prime = \exp(-i\frac{\tau_1}{2}\theta)\alpha^\ast.
\end{equation}
A new type of irreducible decomposition of
\begin{equation}
{\bf 2}_D \otimes {\bf 2}_D =  {\bf 1} \oplus {\bf 3}
\end{equation}
is added to $S_4$, where $\Phi^\dag\Phi$ as {\bf 1} and $\Phi^\dag\Gamma_{1,2,3}\Phi$ as {\bf 3}. 
This is different from ${\bf 2} \otimes {\bf 2} =  {\bf 1} \oplus {\bf 2} \oplus {\bf 1}'$ for vector space.

\section{Parity operator}\label{sec:parityoperator}
We clarify how the mathematical knowledge of $O(3)=SO(3)\times \boldsymbol{Z}_2$ is described in the spinorial space.  To see this, we take $\theta=0$ in Eq.(\ref{Eq:DijO3}), which gives

\begin{eqnarray}
{D_{23}}\left( { - 1,0} \right) &=& \left( {\begin{array}{*{20}{c}}
  0&I \\ 
  I&0 
\end{array}} \right),
\nonumber\\
{D_{31}}\left( { - 1,0 } \right) &=& \left( {\begin{array}{*{20}{c}}
  0&{ - i{\tau _1}} \\ 
  {i{\tau _1}}&0 
\end{array}} \right),
\\
{D_{12}}\left( { - 1,0 } \right) &=& \left( {\begin{array}{*{20}{c}}
  0&{i{\tau _3}} \\ 
  { - i{\tau _3}}&0 
\end{array}} \right).
\nonumber
\end{eqnarray}
Thanks to the phase convention in Eq.(\ref{Eq:SijO3}), in terms of $D_{ij}(1,\theta)$, we find that
\begin{eqnarray}
{D_{23}}\left( { - 1,0} \right) &=& i{D_{31}}\left( { 1,\pi } \right)R,
\nonumber\\
{D_{31}}\left( { - 1,0 } \right) &=& i{D_{12}}\left( { 1,\pi } \right)R,
\label{Eq:ParityOnSpinor}
\\
{D_{12}}\left( { - 1,0 } \right) &=& i{D_{23}}\left( { 1,\pi } \right)R,
\nonumber
\end{eqnarray}
where
\begin{equation}
R = \left( {\begin{array}{*{20}{c}}
  0&{{\tau _2}} \\ 
  {{\tau _2}}&0 
\end{array}} \right).
\end{equation}
These results should be compared with those of $T_{ij}(-1,0)$, which are expressed to be
\begin{eqnarray}
{T_{23}}\left( { - 1,0} \right) &=& {T_{31}}\left( { 1,\pi } \right)P^{(3)},
\nonumber\\
{T_{31}}\left( { - 1,0 } \right) &=& {T_{12}}\left( { 1,\pi } \right)P^{(3)},
\label{Eq:Parity}
\\
{T_{12}}\left( { - 1,0 } \right) &=& {T_{23}}\left( { 1,\pi } \right)P^{(3)},
\nonumber
\end{eqnarray}
where $P^{(3)}$ defined by $P^{(3)} = {\rm diag.}(-1,-1,-1)$ representing the parity inversion generates $\boldsymbol{Z}_2$. 

Comparing Eq.(\ref{Eq:ParityOnSpinor}) with Eq.(\ref{Eq:Parity}), $R$ can be regarded as a parity operator on the $O(3)$ spinor.  In fact, the eigenvalues of $R$ is $(1,1,-1,-1)$; therefore, $R$ is equivalent to $P$ defined by
\begin{equation}
P = \left( {\begin{array}{*{20}{c}}
  I&0 \\ 
  0&{ - I} 
\end{array}} \right).
\end{equation}
The parity operator $P$ that takes $\pm 1$ is linked to the appearance of $\boldsymbol{Z}_2$ in the spinorial space. Furthermore, the phase transformation induced by $\alpha^\prime = e^{-iq}\alpha$, where $q$ is a real number, is described by $\Phi^\prime = e^{-iqP}\Phi$. The parity operator is nothing but the generator of the $U(1)$ transformation. We observe that
\begin{itemize}
	\item $D_{ij}( 1,\pi)$ describes the reflection in the vector space, 
	\item $R$ describes the reflection in the spinor space, leading to the interchange of $\alpha$ and $\alpha^\ast$ and
	\item $P$ describes the generator of the $U(1)$ symmetry.
\end{itemize}
Altogether, it is concluded that our construction of $D_{ij}(-1,\theta)$ is consistent with $O(3)=SO(3)\times \boldsymbol{Z}_2$.

\section{Scenario in particle physics}\label{sec:scenario}
\begin{table}[b]
\caption{\label{table1}Transformation property of  the new fields ($\nu_{Ri}, N, \phi$), lepton doublet $L$ and Higgs doublet $H$ under $S_4$, $SU(2)_L$ and $U(1)_Y$ .}
\begin{indented}
\item[]\begin{tabular}{@{}ccccccc}
\br
& Scenario A &Scenario B &  & &  &  \\
& $\nu_{R12}$,  $\nu_{R3}$ &$\vec{\nu}_R$& $N$ & $\phi$ & $\vec{L}$ & $\vec{H}$  \\
\mr
$S_4$ & ${\bf 2}$, ${\bf 1}$ & ${\bf 3}$ & ${\bf 2}_D$ & ${\bf 2}_D$ & ${\bf 3}$ & ${\bf 3}$  \\
$SU(2)_L$ & ${\bf 1}$, ${\bf 1}$ & ${\bf 1}$ & ${\bf 1}$ & ${\bf 1}$ & ${\bf 2}$ & ${\bf 2}$  \\
$U(1)_Y$ & $0$, $0$ & $0$ & $0$ & $0$ & $-1/2$ & $1/2$  \\
\br
\end{tabular}
\end{indented}
\end{table}

Although our main goal of this paper is to construct the spinor representation of the $S_4$ group, it is plausible to present brief discussions on possible phenomenological application of doublet representation ${\bf 2}_D$ to particle physics. The application is very limited because our spinor $\Phi$ involves the complex number and its complex conjugate, namely, particle and antiparticle. Since physical quantum numbers are placed on $\Phi$, $\Phi$ is singlet at least in the standard model. For a Dirac spinor $\psi$, a fermionic doublet should be composed of $\psi$ and $\psi^C$ instead of $\psi^\ast$ to be consistent with the Dirac spinor,  where $\psi^C$ is the charge-conjugated state of $\psi$ \cite{KitabayashiYasue}. Since $\psi^C\neq\psi$ should be satisfiled to form the $O(3)$ spinor, a Majorana fermion satisfying that $\psi^C=\psi$ cannot be accepted. Furthermore, our fermionic $O(3)$ spinor is required to be vectorlike to ensure that the $O(3)$ transformation properly works. Let $\Psi=(\psi_1,\psi_2)^T$ be an $SO(3)$ spinor, then the $O(3)$ transformation yields $\Psi^\prime=S_{ij}(-1,\theta)\Psi^C$, which cannot be satisfied if $\psi_a$ ($a=1,2$) are chiral because the left-handed $\psi_a$ gives the right-handed $\psi^C_a$ or vice versa; therefore, $\psi_a$ is set to be vectorlike. Since the quantum number of the neutral $\psi_a$ is given by the $U(1)$ charge associated with $P$, which is proportional to 1 for $\psi_a$ and $-1$ for $\psi^C_a$, the condition of $\psi^C_a \neq \psi_a$ is naturally satisfied.  

Since interactions with $O(3)$ spinors are allowed for the standard model particles carrying no quantum numbers associated with the standard model symmetries, possible candidates to participate in our $O(3)$ model are the right-handed neutrinos, which can couple to the left-handed leptons. Let us start by adding the following new fields to the particle contents of the standard model:
\begin{itemize}
\item Three right-handed neutrinos: $\nu_{Ri}$ $(i =1,2,3)$,  
\item Vectorlike neutral fermion: $N^0=\left(
  \begin{array}{c}
     N^0_1\\
     N^0_2\\
  \end{array}
\right)$,
\item Neutral scalar: $\phi^0=\left(
  \begin{array}{c}
     \phi^0_1\\
     \phi^0_2\\
  \end{array}
\right)$,
\end{itemize}
where $N^0$ and $\phi^0$ are spinors of $SO(3)$. Suppose that these additional fields are assigned to one of the irreducible representations of $S_4$ as follows:
\begin{itemize}
\item For three right-handed neutrinos, $\nu_{R1,R2}$ and $\nu_{R3}$, respectively, transform as {\bf 2} and {\bf 1} (Scenario A),
\begin{equation}
\nu_{R12}=(\nu_{R1},\nu_{R2}) : {\bf 2}({\bf 1},0),
\quad
\nu_{R3} : {\bf 1}({\bf 1},0),
\end{equation}
or $\nu_{R1,R2,R3}$ transform as {\bf 3} (Scenario B),
\begin{equation}
\vec{\nu}_R : {\bf 3}({\bf 1},0)
\end{equation}
where $\vec{\nu}_R = \nu_{R1} \vec{i} +  \nu_{R2} \vec{i} +  \nu_{R3} \vec{i}$;
\item For $N^0$ and $\phi^0$, both transform as ${\bf 2}_D$ described by 
\begin{equation}
N=\left(
  \begin{array}{c}
     N^0   \\
     N^{0C}\\
  \end{array}
\right),
\phi=\left(
  \begin{array}{c}
     \phi^0   \\
     \phi^{0\ast}\\
  \end{array}
\right) :  {\bf 2}_D({\bf 1},0), 
\end{equation}
where $N^{0C}$ denotes the charge conjugated state;
\item For the lepton doublet $L$ in the standard model, $L$ is assigned to {\bf 3} represented by $\vec{L}$:
\begin{equation}
\vec{L} : {\bf 3}({\bf 2},-1/2),
\end{equation}
\item For the Higgs $H$ in the standard model, $H$ is accompanied by two extra Higgses to form {\bf 3} represented by $\vec{H}$:
\begin{equation}
\vec{H} :  {\bf 3}({\bf 2},1/2).
\end{equation}
\end{itemize}
The representations of $S_4$, $SU(2)_L$ and $U(1)_Y$ are shown as $S_4 (SU(2)_L, U(1)_Y)$.  These assignments are listed in Table \ref{table1}.

In this setup, relevant $S_4$ singlet terms related to $\vec{L}$ and $\nu_R$'s are $\bar{\vec{L}} \vec{H} \nu_{R12}$, $\bar{\vec{L}} \vec{H} \vec{\nu}_R$ and $\bar{\vec{L}} \vec{H} \nu_{R3}$, which are allowed because of ${\bf 3} \otimes {\bf 3} ={\bf 1} \oplus {\bf 3} \oplus {\bf 2}\oplus {\bf 3'}$. Moreover, $S_4$ singlet terms related to the $O(3)$ spinor $N$ are ${\bar{\nu}}_{R3} \phi^{\dag} N$ and $\bar{\vec{\nu}}_R \phi^{\dag} \vec{\Gamma}N$, which are allowed because of ${\bf 2}_D \otimes {\bf 2}_D =  {\bf 1} \oplus {\bf 3}$. The $S_4$ invariant Lagrangian is given by
\begin{eqnarray}
\mathcal{L}_A &=& \lambda_{1}(\bar{\vec{L}} \vec{H} \nu_{R12})_{{\bf 3}\otimes{\bf 3}\otimes{\bf 2}} + \lambda_{2}(\bar{\vec{L}} \vec{H} \nu_{R3})_{{\bf 3}\otimes{\bf 3}\otimes{\bf 1}}  \nonumber \\
&& + \lambda_{3} (\bar{\nu}_{R3} \phi^{\dag} N) + h.c., 
\label{Eq:Lagrangian_nu_2_1}
\end{eqnarray}
for the Scenario A, where $\phi^{\dag} N$ is {\bf 1} and three tensor products placed in the subscripts are so chosen to form singlet products, or by
\begin{eqnarray}
\mathcal{L}_B &=& \lambda_{1}(\bar{\vec{L}} \vec{H} \vec{\nu}_R)_{{\bf 3}\otimes{\bf 3}\otimes{\bf 3}}
+ \lambda_{2} (\bar{\vec{\nu}}_R \phi^{\dag}\vec{\Gamma} N) + h.c.,
\label{Eq:Lagrangian_nu_3}
\end{eqnarray}
for the Scenario B, where $\phi^{\dag}\vec{\Gamma} N$ is {\bf 3}. 

These particle models may have a potential ability to solve problems in particle physics as well as in cosmology. For example, in the  model of the Scenario A based on $\mathcal{L}_A$, one right-handed neutrino $\nu_{R3}$ and other two right-handed neutrinos $\nu_{R1},\nu_{R2}$ are distinguished. This difference may give some hints about the problem of the neutrino mass hierarchy. Moreover, additional neutral fermion $N^0$ as well as neutral scalar $\phi^0$ may be candidate of dark matter \cite{DarkMatter}. Since only $\nu_{R3}$ can interact with $N^0$ and $\phi^0$, it suggests that the dark matter sector and neutrino sector are closely related to each other.

\section{Summary}\label{sec:summary}
We have succeeded to describe $O(3)$ rotations acting on the four-component spinor $\Phi$, which contains the $SO(3)$ spinor $\alpha$ and its complex conjugate $\alpha^\ast$: $\Phi=(\alpha, \alpha^\ast)^T$.  The four-component spinor behaves as a spinor doublet of $O(3)$, ${\bf 2}_D$. The extended $4 \times 4$ Pauli matrices can be given by the block diagonal form of $\Gamma_i = {\rm diag.}(\tau_i,\tau^\ast_i)$ ($i$=1,2,3). The transformation matrices describing the reflection consist of the well-known $SO(3)$ rotation matrices $S_{ij}(1,\theta)$ and are determined to be: $S_{23}(-1,\theta)= S_{23}(1,\theta)$, $S_{31}(-1,\theta)= -iS_{31}(1,\theta)\tau_1$ and $S_{12}(-1,\theta)= iS_{12}(1,\theta)\tau_3$. We find that $D_{ij}(\sigma,\theta)$ acting on $\Phi$ take the form:
\begin{eqnarray}
D_{ij}(1,\theta)&=&\left(
  \begin{array}{cc}
     S_{ij}(1,\theta) & 0 \\
     0 & S_{ij}^\ast (1,\theta)  \\
  \end{array}
\right),
\nonumber\\
D_{ij}(-1, \theta)&=&\left(
  \begin{array}{cc}
     0&  S_{ij}(-1,\theta)  \\
     S^\ast_{ij}(-1,\theta)&0  \\
  \end{array}
\right).
\end{eqnarray}
The effect of the reflection on $\Phi$ dictates from the transformation of the $SO(3)$ spinor $\alpha$. For example, under the $2$-$3$ rotation with $\sigma =-1$, $\alpha$ gets transformed according to
\begin{eqnarray}
&&\alpha^\prime = \exp(-i\frac{\tau_1}{2}\theta)\alpha~(\sigma=1),
\nonumber\\
&&\alpha^\prime = \exp(-i\frac{\tau_1}{2}\theta)\alpha^\ast~(\sigma=-1).
\end{eqnarray}
The bilinear products of $\Phi^\dag\Phi$ and $\Phi^\dag\vec{\Gamma}\Phi$, respectively, behave as {\bf 1} and {\bf 3} of $O(3)$, which in turn provide ${\bf 2}_D \otimes {\bf 2}_D =  {\bf 1} \oplus {\bf 3}$. 

It is also demonstrated that the $\boldsymbol{Z}_2$ parity in $O(3)=SO(3)\times \boldsymbol{Z}_2$ is represented by the off-diagonal parity operator $R$ on the $O(3)$ spinor inducing the mixing of $\alpha$ and $\alpha^\ast$. This off-diagonal parity operator is equivalent to the diagonal parity operator $P$ whose eigenvalues are $\pm 1$, which is responsible for the $\boldsymbol{Z}_2$ parity. In a physical language, $P$ is the generator of the $U(1)$ symmetry, which provides $q$ as an $U(1)$ charge of $\Phi$ associated with $\Phi^\prime = e^{-iqP}\Phi$.

The application of the doublet representation ${\bf 2}_D$ to particle physics is very restricted owing to the inherent property of $\Phi$ containing both a particle and an antiparticle. The $O(3)$-doublet spinors should be singlets in the standard model. In Sec.\ref{sec:scenario}, we have presented models of vectorlike particles as fermions and neutral scalars belonging to ${\bf 2}_D$. Since new particles as ${\bf 2}_D$ must be neutral, this restriction may be good news for cosmological dark matter, which is considered to be neutral. Detailed study of effects of ${\bf 2}_D$ on particle physics including dark matter scenario \cite{KitabayashiYasue} as well as our $S_4$ scenarios discussed in Sec \ref{sec:scenario} are left for our future study.

\end{document}